\documentclass[a4paper,11pt]{article}
\pdfoutput=1 

\usepackage{jheppub} 

\usepackage[T1]{fontenc} 

\usepackage{float}
\usepackage{caption}
\usepackage{subcaption}

\title{\boldmath Impact of the Hubble tension on the $r-n_s$ contour}

\author[a,b]{Jun-Qian Jiang,
}
\author[c]{Gen Ye,}
\author[a,b,d,e]{Yun-Song Piao}

\affiliation[a]{School of Fundamental Physics and Mathematical
   Sciences, Hangzhou Institute for Advanced Study, UCAS, Hangzhou
    310024, China}

\affiliation[b]{School of Physics, University of Chinese Academy
of Sciences, Beijing 100049, China}

\affiliation[c]{Leiden University, Instituut-Lorentz for Theoretical Physics, 2333CA, Leiden, Netherlands}

\affiliation[d]{International Center for Theoretical Physics
    Asia-Pacific, Beijing/Hangzhou, China}

\affiliation[e]{Institute of Theoretical Physics, Chinese
    Academy of Sciences, P.O. Box 2735, Beijing 100190, China}

\emailAdd{jiangjq2000@gmail.com}

\emailAdd{ye@lorentz.leidenuniv.nl}

\emailAdd{yspiao@ucas.ac.cn}

\abstract{The injection of early dark energy (EDE) before the
recombination, a possible resolution of the Hubble tension, will
not only shift the scalar spectral index $n_s$ towards $n_s=1$,
but also be likely to tighten the current upper limit on
tensor-to-scalar ratio $r$. In this work, with the latest CMB
datasets (Planck PR4, ACT, SPT and BICEP/Keck), as well as BAO and SN, we
confirm this result, and discuss its implication on inflation. We
also show that if we happen to live with EDE, how the different
inflation models currently allowed would be distinguished by
planned CMB observations, such as CMB-S4 and LiteBIRD. }

\begin{document}
\maketitle
\flushbottom

\section{Introduction}
\label{sec:intro}

Inflation is a current paradigm of the very early universe,
predicting a nearly scale-invariant primordial scalar perturbation
and primordial gravitational wave (GW). A combined analysis of the
CMB observations by Planck with other datasets showed the scalar
spectral index $n_s = 0.965\pm 0.004$ (68\% CL)
\cite{Planck:2018vyg}, while its combination with the recent
BICEP/Keck dataset showed tensor-to-scalar ratio $r<0.036$ (95\% CL)
\cite{BICEP:2021xfz}. However, both results are based on the
$\Lambda$CDM model, and cosmological model-dependent.

Recently, some inconsistencies in cosmological observations have
been suggested, see e.g.
Refs.\cite{Perivolaropoulos:2021jda,Abdalla:2022yfr} for recent
reviews. The most well-known is the Hubble tension, which have
inspired the exploring beyond $\Lambda$CDM model,
e.g.\cite{DiValentino:2019qzk,Handley:2019tkm,DiValentino:2016hlg,Mortsell:2018mfj,Vagnozzi:2019ezj,Knox:2019rjx,DiValentino:2019jae,Schoneberg:2021qvd}.
In the early dark energy (EDE) resolution
\cite{Karwal:2016vyq,Poulin:2018cxd} of the Hubble tension, a
unknown EDE component before the recombination lowered the sound
horizon $r_s=\int {c_s\over H(z)}dz$,
where $c_s$ and $H(z)$ is the sound speed and Hubble parameter before the recombination respectively,
so that $H_0$ is lifted,
which, however, also brings an unforeseen effect on our
understanding on inflation.

It has been found that the injection of EDE \footnote{Actually,
``EDE" corresponds to EDE+$\Lambda$CDM, which is only a
pre-recombination modification to $\Lambda$CDM, and the evolution
after the recombination must still be $\Lambda$CDM-like.} will
alter the results of $n_s$
\cite{Ye:2021nej,Jiang:2022uyg,Smith:2022hwi,Jiang:2022qlj,Cruz:2022oqk}, specially if
$H_0\gtrsim 72$ km/s/Mpc, $n_s$ will be shifted to $n_s=1$
($|n_s-1|\sim {\cal O}(0.001)$)\cite{Ye:2020btb}, and the current
limit on tensor-to-scalar ratio $r$ \cite{Ye:2022afu}, see also
\cite{DiValentino:2018zjj,Giare:2022rvg,Calderon:2023obf} for
studies on the possibilities of $n_s=1$. Thus some inflation
models that had been considered possible in the $\Lambda$CDM model
might be excluded, while some inflation models which had been
excluded might be reconsidered as possible. However, it is
interesting to ask whether the result on $r-n_s$ is still credible
with the Planck PR4 \cite{Planck:2020olo}, as well as latest
ACT \cite{ACT:2020frw}, SPT-3G \cite{SPT-3G:2021eoc,SPT-3G:2022hvq}
dataset. In this work, we will investigate this issue.

It is expected that cosmological observations would have the
ability to discriminate between different inflation models if
their precision becomes high enough. In upcoming decade, the
Simons Observatory\cite{SimonsObservatory:2018koc},
CMB-S4\cite{CMB-S4:2016ple}, as well as the LiteBIRD
satellite\cite{LiteBIRD:2020khw}, will play significant roles in
improving the constraint on $r-n_s$. The combination of CMB-S4 and
LiteBIRD will be able to reach $\sigma(n_s) \sim 0.005$ and
$\sigma(r) < 10^{-3}$.
{ 
This forecast is obtained assuming AdS-EDE, in which the potential
is $\phi^4$-like but with an anti-de Sitter (AdS) phase
\cite{Ye:2020btb}, but we expect this should be similar for other EDE models.
}
Here, we also will show how different
inflation models allowed by the present observations can be
distinguished by upcoming CMB-S4 and LiteBIRD experiments.

The outline of paper is as follows. We show in
\autoref{sec:cur_constraint} the impact of EDE on the current
constraint on $r-n_s$ and its implication for inflation. The
abilities of CMB-S4 and LiteBIRD with EDE is forecasted in
\autoref{sec:fur_constraint}. We conclude in
\autoref{sec:conclusion}.
The priors for all cosmological parameters employed in our analysis are shown in \autoref{sec:priors}.
In \autoref{sec:appendix}, we present
the results with the tensor spectral index $n_t$ and running of scalar spectral index $\alpha_s$.
The details on noise power spectrum and delensing used in our forecast are shown in \autoref{sec:appendixB}
and posterior distributions for all cosmological parameters are shown in \autoref{sec:all}.

\section{Results with current data}
\label{sec:cur_constraint}

\subsection{Datasets and models}
Datasets are as follows:
\begin{itemize}
\item \textbf{PR4}: The latest release of Planck maps (PR4), with
the NPIPE code \cite{Planck:2020olo}.
We take use of the \texttt{hillipop} likelihood
\cite{Couchot:2016vaq} for high-$\ell$ part and \texttt{lollipop}
\cite{Tristram:2020wbi} as low-$\ell$ polarization likelihood. As
for the low-$\ell$ TT power spectrum, we take use of the public
\texttt{commander} likelihood \cite{Planck:2019nip}. Planck PR4
lensing likelihood \cite{Carron:2022eyg} is included. \item
\textbf{ACT}: The ACTPol Data Release 4 (DR4) \cite{ACT:2020frw}
likelihood for all TT, TE, EE power spectrum, which has already
been marginalized over SZ and foreground emission. \item
\textbf{SPT}: The SPT-3G Y1 data \cite{SPT-3G:2021eoc} of the TE,
EE power spectrum and the recent TT power spectrum data
\cite{SPT-3G:2022hvq}. We take use of the likelihoods adapted for
\texttt{cobaya}\footnote{\url{https://github.com/xgarrido/spt_likelihoods}}.
\item \textbf{BK18}: The latest BICEP/Keck likelihood on the BB
power spectrum\cite{BICEP:2021xfz}. \item \textbf{BAO}: The 6dF
Galaxy Survey \cite{Beutler:2011hx} and SDSS DR7 main Galaxy
sample \cite{Ross:2014qpa} for the low-$z$ part. The eBOSS DR16
data \cite{eBOSS:2020yzd}, which include LRG, ELG, Quasar,
Ly$\alpha$ auto-correlation and Ly$\alpha$-Quasar
cross-correlation, for the high-$z$ part
\footnote{ 
Although it has been argued that there may be some inconsistency with the Lyman-$\alpha$ BAO data (see e.g. Ref.\cite{Cuceu:2019for}), the joint constraint does not depend on whether the Lyman-$\alpha$ BAO data is included or not in the new data \cite{Schoneberg:2022ggi}. And we have checked that our conclusion does not depend on the selection of BAO data.
}
. We use a combined
likelihood with the BOSS DR12 BAO data \cite{BOSS:2016wmc}. \item
\textbf{SN}: The uncalibrated measurement of Pantheon+ on the Type
Ia supernovae (SNe Ia) ranging in redshift from $z= 0.001$ to
2.26. \cite{Brout:2022vxf}
\end{itemize}
In this work, for the CMB observations, we consider the
combination of Planck and BICEP/Keck, and also the combination of
Planck, ACT, SPT and BICEP/Keck. { 
When the small scales of CMB spectrum can be complemented by ground-based CMB observations focused on small scales,
we cut the PR4 TT spectrum to $\ell$<1000
as there are some doubts about the Planck TT high-$\ell$ part (e.g. \cite{Addison:2015wyg, Planck:2016tof, Motloch:2019gux}).
Meanwhile, it has been shown that EDE seems to resolve Hubble tension in this scenario.
{ 
As we are investigating the impact of Hubble tension, we only focus on the scenario where Hubble tension can be resolved.
}
}
Besides, we cut the \texttt{hillipop} EE
likelihood to $\ell>150$ in order to avoid the correlations with
the \texttt{lollipop} likelihood. BAO and SN, which do not
conflict with the CMB observations, are included in all datasets.

{ 
In canonical scalar field models of EDE,
the EDE field is frozen initially due to the Hubble friction,
and thus contributes extra energy before recombination,
resulting in a lower sound horizon $r^*_\text{s}$,
so a higher $H_0$, as
$\theta_\text{s}=r_\text{s}/D_A\sim r_\text{s}H_0$ is set
precisely by CMB observations.
As the expansion of the Universe slows down, the EDE begins to roll down and lose its energy.
The requirement that EDE must decay
fast enough to avoid disruption of the CMB fit has motivated
different EDE models.
Here, we consider the axion-like EDE
\cite{Poulin:2018cxd}, which is achieved by a scalar field with
axion-like potential (see recent
\cite{McDonough:2022pku,Cicoli:2023qri} for models in string
theory),
and the AdS-EDE
\cite{Ye:2020btb,Ye:2020oix,Jiang:2021bab}.
In the axion-like EDE model, the fast decay is achieved through an oscillation phase,
which has an equation-of-state parameter $w=1/2$ (here we consider the case $n=3$ for the axion-like EDE potential).
While in the AdS-EDE model, it is achieved through an AdS phase, which has an equation-of-state parameter $w>1$.
Here we consider an AdS-EDE with a $\phi^4$-like potential.
In both cases, EDE can decay faster than the radiation.
In our analysis, we use the phenomenological parameter
$z_c$, which means the redshift when the field starts rolling,
and $f_\text{EDE}$, which means the energy fraction of EDE at $z_c$,
instead of the theoretical parameters.
}


The MCMC sampling is performed using \texttt{Cobaya}
\cite{Torrado:2020dgo}, while we use the modified \texttt{CLASS}
\cite{Blas:2011rf} \footnote{The codes are available at
\url{https://github.com/PoulinV/AxiCLASS} for axion-like EDE and
\url{https://github.com/genye00/class\_multiscf} for AdS-EDE.} to
calculate models.
{ 
The cosmological parameters include the six standard $\Lambda$CDM parameters
$\{H_0, n_s, \omega_b, \omega_\text{cdm}, \tau_\text{reio}, A_s\}$,
where $H_0$ is Hubble constant, $n_s$ is scalar spectral index,
$\omega_b = \Omega_b h^2 \, (h = H_0 / 100)$ is baryon density today,
$\omega_\text{cdm} = \Omega_\text{cdm} h^2$ is dark matter density today,
$\tau_\text{reio}$ is optical depth,
$A_s$ is the primordial curvature perturbations.
In addition to these,
we also sample on
the tensor/scalar ratio
$r={A_T\over A_s}$ at the pivot scale 0.05 Mpc$^{-1}$ and 
the parameters of EDE models $z_c$ and $f_\text{EDE}$.
For axion-like EDE, we also sample the initial phase $\Theta_\text{ini}$.
}

\subsection{Results}

\begin{figure}[h]
    \centering
    \includegraphics[width=0.85\textwidth]{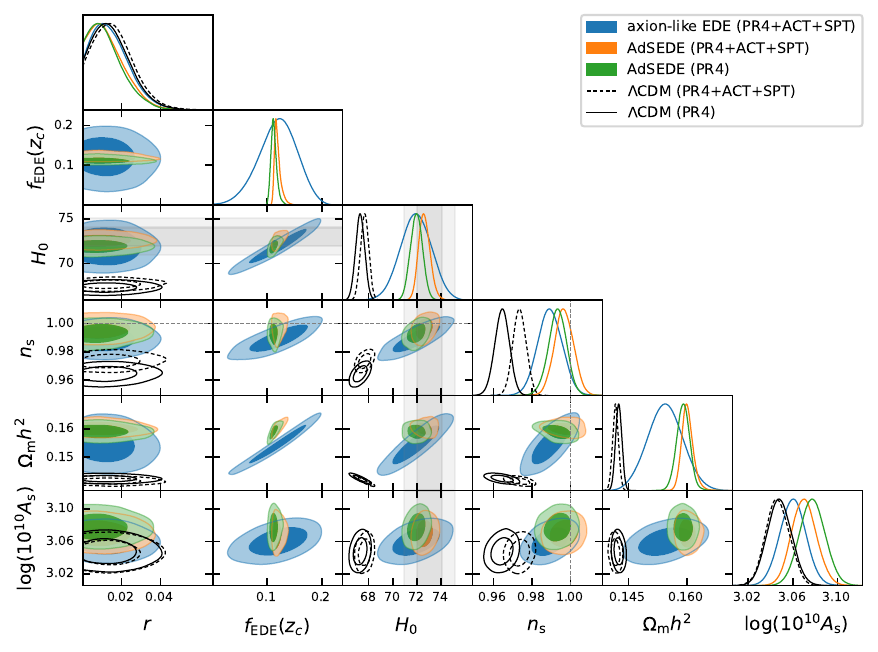}
\caption{Marginalized posterior distributions (68\% and 95\%
confidence intervals) for relevant parameters in different
models with different datasets. See \autoref{sec:all} for all parameters. BK18 is included in all datasets.
Constraint of R21\cite{Riess:2021jrx} on $H_0$ is shown as a gray band.}
    \label{fig:EDE_constraints}
\end{figure}

\begin{table}[h]
    \centering
    \scriptsize
    \begin{tabular}{c|ccc} \hline \hline
parameters & AdSEDE (PR4+ACT+SPT) & axion-like EDE (PR4+ACT+SPT) & AdSEDE (PR4) \\ \hline
$\log_{10}(z_c)            $ &  $3.470(3.4647)^{+0.022}_{-0.027}$&  $3.514(3.5022)\pm 0.044    $&  $3.540(3.4536)^{+0.035}_{-0.041}$\\
$f_\text{EDE}              $ &  $0.1193(0.11326)^{+0.0028}_{-0.0070}$&  $0.119(0.0967)^{+0.038}_{-0.032}$&  $0.1123(0.1225)^{+0.0040}_{-0.0062}$\\
$\Theta_\text{ini}         $ & & $2.737(2.752)\pm 0.087     $ & \\
$r              $ &  $< 0.0303          $&  $< 0.0324$&  $<0.0284          $\\
$\log(10^{10} A_\mathrm{s})$ &  $3.070(3.0663)\pm 0.011    $&  $3.060(3.0613)\pm 0.012    $&  $3.078(3.0706)\pm 0.011    $\\
$n_\mathrm{s}   $ &  $0.9966(0.99438)\pm 0.0049 $&  $0.9893(0.9857)\pm 0.0063  $&  $0.9932(0.9903)^{+0.0049}_{-0.0044}$\\
$H_0            $ &  $72.62(72.255)^{+0.42}_{-0.51}$&  $71.9(71.12)\pm 1.3        $&  $71.96(71.58)\pm 0.50      $\\
$\Omega_\mathrm{b} h^2$ &  $0.02300(0.022928)\pm 0.00016$&  $0.02251(0.022424)^{+0.00015}_{-0.00017}$&  $0.02321(0.023108)^{+0.00018}_{-0.00016}$\\
$\Omega_\mathrm{c} h^2$ &  $0.1365(0.13588)^{+0.0013}_{-0.0016}$&  $0.1319(0.12895)\pm 0.0045 $&  $0.1352(0.13622)\pm 0.0018 $\\
$\tau_\mathrm{reio}$ &  $0.0521(0.0505)\pm 0.0060  $&  $0.0542(0.0561)\pm 0.0056  $&  $0.0591(0.0580)\pm 0.0063  $\\
\hline
$\Omega_\mathrm{m}         $ &  $0.3038(0.30541)\pm 0.0050 $&  $0.2987(0.29929)\pm 0.0050 $&  $0.3072(0.3122)\pm 0.0053  $\\
$S_8                       $ &  $0.868(0.8676)\pm 0.010    $&  $0.848(0.8448)\pm 0.012    $&  $0.871(0.8751)\pm 0.011    $\\ \hline
$\chi^2 - \chi^2_\text{$\Lambda$CDM}$ & $-6.0$              &  $-11.2$                        &  $2.0$ \\
\hline \hline
    \end{tabular}
\caption{68\% confidence intervals for the parameters in different
EDE models with different datasets, the best-fit values are shown
in parentheses. And for the one-tailed distribution we show 95\%
confidence intervals. BK18 is included in all datasets.}
    \label{tab:EDE_constraints}
\end{table}

Our results are shown in \autoref{fig:EDE_constraints} and
\autoref{tab:EDE_constraints}.
{ 
Since we are exploring the impact on the $r$-$n_s$ contour of EDE models in light of the Hubble tension,
we choose the models and corresponding data set that allow enough EDE to resolve the tension,
even it cannot degenerate to $\Lambda$CDM.
}
In such EDE-like models
\footnote{
{ 
In the AdS-EDE model, with only PR4+BAO+SN(+BK18) dataset,
$H_0\gtrsim 72$km/s/Mpc is acquired due to the existence of the AdS bound,
where the field failed to climb out of the AdS potential for
small $f_\text{EDE}$.
Therefore, the posterior will not be connected to the $\Lambda$CDM even if the parameter prior range allows it.
However,
}
$\Lambda$CDM is allowed in some AdS-EDE models in which the depth of AdS vacuum is varied and regarded as a MCMC parameter,
while AdS-EDE and higher $H_0$ are still preferred, see e.g. \cite{Ye:2020oix}.
Here we take AdS-EDE with $\phi^4$-like potential for simplicity.
}, we have $H_0\gtrsim 72$km/s/Mpc, compatible with the
recent local $H_0$ measurement \cite{Riess:2021jrx} (hereafter
R21), see also
Refs.\cite{LaPosta:2021pgm,Smith:2022hwi,Jiang:2022uyg} with
Planck PR3, and \autoref{sec:appendix} for the results with $n_t$ and the
running $\alpha_s$ of $n_s$.

\begin{figure}[h]
    \centering
    \includegraphics[width=0.62\textwidth]{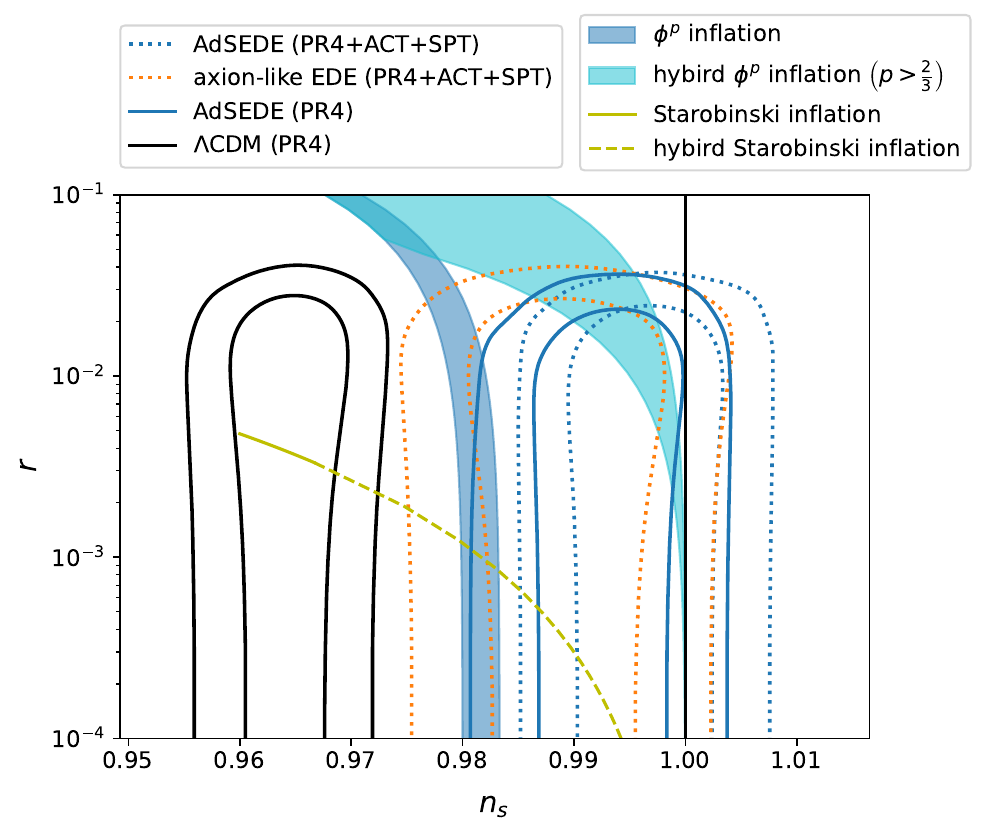}
\caption{The $r$-$n_s$ contour for different models and different
combinations of datasets. We also plot the predictions of some
inflation models on $n_s$ and $r$. The yellow line and the blue
band represent Starobinski inflation \cite{Starobinsky:1980te} and
$\phi^p$ inflation, respectively, with $50\leqslant N_* \leqslant
60$. The dashed yellow line and the cyan band correspond to their
hybrid variants where the inflation ended in a deep slow-roll
region so that $N_*\gg 60$. The left bound ($p=2/3$) of the cyan
band is associated with the monodromy inflation
\cite{Silverstein:2008sg,McAllister:2008hb} and the right bound
($p \rightarrow \infty$) is associated with the power-law
inflation \cite{Abbott:1984fp}.}
    \label{fig:ns_r}
\end{figure}

Apart from the uplift of $H_0$, another dramatic change is the
shift of the $n_s$-$r$ contour, which is shown in
\autoref{fig:ns_r}, specially $n_s$ is shifted close to $n_s=1$
\footnote{It has even been found that if the Hubble tension is
fully resolved in such EDE models, it may suggest a
Harrison-Zeldovich spectrum (i.e. $n_s=1$) \cite{Jiang:2022qlj}.}.
The shift of $n_s$ is a common result of any prerecombination
resolution (without modifying the recombination physics) for the
Hubble tension \cite{Ye:2021nej}. The shift of $n_s$ with respect
to $H_0$ can be approximated as \cite{Ye:2021nej,Jiang:2022uyg}:
\begin{equation}
\delta n_s \approx 0.4 \frac{\delta H_0}{H_0}.
\end{equation}
The upper limit of $r$ (e.g. $r<0.028$ in AdS-EDE) has been
slightly lower than that under the $\Lambda$CDM model $r<0.330$ (PR4)
\footnote{See \autoref{tab:LCDM_constraints} for detailed values. Our results differ slightly from
Ref.\cite{Tristram:2021tvh} due to the different CMB data
combinations selected and the BAO+SN dataset.}
and $r< 0.0352$ (PR4+ACT+SPT)
This is mainly
because the slight uplifts of $\omega_m$ and $A_s$ in EDE models,
compared with $\Lambda$CDM, enhance the lensing spectrum between
$200 < \ell < 800$.
{
The constraint on $r$ mainly comes from the observation of the CMB BB spectrum,
where the effect of lensing is the main contributor in the present observational range.
A larger lensing spectrum means a higher contribution,
which in other words is smaller $r$ after subtracting this contribution \cite{Ye:2022afu}.
}

In well-known single field slow-roll inflation models, $n_s$
follows
\cite{Mukhanov:2013tua,Kallosh:2013hoa,Roest:2013fha,Martin:2013tda}
\begin{equation}
    n_s-1 = -{{\cal O}(1)\over N_*} \label{ns}
\end{equation}
in large $N_*$ limit,
where $N_*$ is the e-folding number spent during inflation
($M_p=1$) \begin{equation}
N_*\approx\Delta N+\int_{\phi_e}^{\phi_c}{d\phi\over
\sqrt{2\epsilon}}=
\left(\int_{\phi_c}^{\phi_*}+\int_{\phi_e}^{\phi_c}\right){d\phi\over
\sqrt{2\epsilon}}\approx N(\phi_*), \label{N}\end{equation}
{ 
where $\epsilon = -\dot{H}/H^2$ is the slow-roll parameter
}
and $\phi_*$ is the value of the field at which the perturbation mode with $k=k_*$ exits
horizon, which sets $N_*$.
Here $\phi_c$ is the value of the field when $\Delta N\approx 60$ was reached
and $\phi_e$ is the value of the field when inflation ended in these models.
Thus both $n_s$ and $r$ are related to $N_*$ rather than
$\Delta N$. It is usually thought that inflation ended at
$\phi_e$ when the slow-roll condition breaks down, we have $N_*=\Delta N\approx 60$ (inflation ends around
$\sim10^{15}\text{GeV}$). However, if inflation ended prematurely
at $\phi_c$ during the slow-roll regime when $\epsilon\ll 1$, we will have $N_*\gg 60$ but
still $\Delta N\approx 60$ \cite{Kallosh:2022ggf,Ye:2022efx}.

It is interestingly found that some inflation models, such as the
power-law inflation \cite{Abbott:1984fp} and the $\phi^p$
inflation
\cite{Linde:1983gd,Silverstein:2008sg,McAllister:2008hb,Kaloper:2008fb},
which were disfavored in the $\Lambda$CDM model, are now compatible
with the EDE, see \autoref{fig:ns_r}. In such inflation models,
the inflation might end up by a waterfall instability, which is
similar to the hybrid inflation \cite{Linde:1991km,Linde:1993cn},
at a deep slow-roll region $\epsilon\ll 1$, so that $N_*\gg 60$.
The perturbation modes near $N_*$ can be just at CMB band, so we
have $|n_s-1|\ll {\cal O}(0.01)$.
See also e.g. \cite{DAmico:2020euu,DAmico:2021vka,Takahashi:2021bti,DAmico:2021fhz} for recent significant endeavors in inflation model with $n_s=1$.

\section{Forecast with CMB-S4 and LiteBIRD}
\label{sec:fur_constraint}

It is also significant to investigate the impact of EDE on the
constraining power of CMB-S4 and LiteBIRD. The CMB-S4
\cite{Abazajian:2019eic} will cover the sky area of
$f_\text{sky}=0.4$, so $20<\ell<5000$ for CMB, while the LiteBIRD
\cite{LiteBIRD:2020khw} is a satellite covering a larger sky area.
Here, we assume $f_\text{sky}=0.9$ (so $2<\ell<200$) for LiteBIRD,
and also set $\ell_\text{min}=200$ for CMB-S4 to avoid the
correlation between them. And relevant noise power spectrum and
delensing are presented in \autoref{sec:appendixB}.

We use the Fisher matrix to make predictions.
{ 
The Fisher matrix is
the expectation of the Hessian of the log-likelihood \cite{Jungman:1995av}:
\begin{equation} \label{eq:Fisher}
    F_{i j}=\left\langle H_{i j}\right\rangle = \left\langle \frac{\partial^2 \ln \mathcal{L}}{\partial \theta_i \partial \theta_j} \right\rangle = \sum_\ell \operatorname{Tr}\left\{ \frac{\partial C_\ell}{\partial \theta_i} \mathcal{C}_\ell^{-1} \frac{\partial C_\ell}{\partial \theta_j} \mathcal{C}_\ell^{-1}\right\}
\end{equation}
where $\mathcal{C}_\ell^{X Y} \equiv \left(\frac{2\ell+1}{2} f_\text{sky}\right)^{-1/2} \left( C_{\ell}^{X Y}+N_{\ell} \delta^{X
Y} \right)$ is the covariance matrix. $X, Y$ represent T, E, B respectively,
$f_\text{sky}$ is the fraction of the sky,
and $C_\ell$, $N_\ell$ is the power spectrum and noise curve (defined in \autoref{sec:appendixB}) respectively.
Thus we can estimate the parameter probability covariance matrix:
$\mathbf{C} = [F]^{-1}$.
}

\begin{figure}[h]
    \centering
    \begin{subfigure}[b]{0.49\textwidth}
         \centering
         \includegraphics[width=\textwidth]{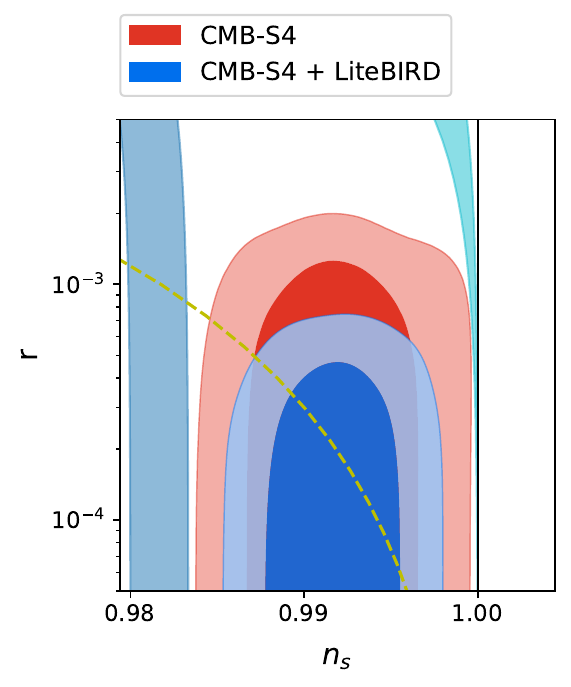}
    \end{subfigure}
    \begin{subfigure}[b]{0.50\textwidth}
         \centering
         \includegraphics[width=\textwidth]{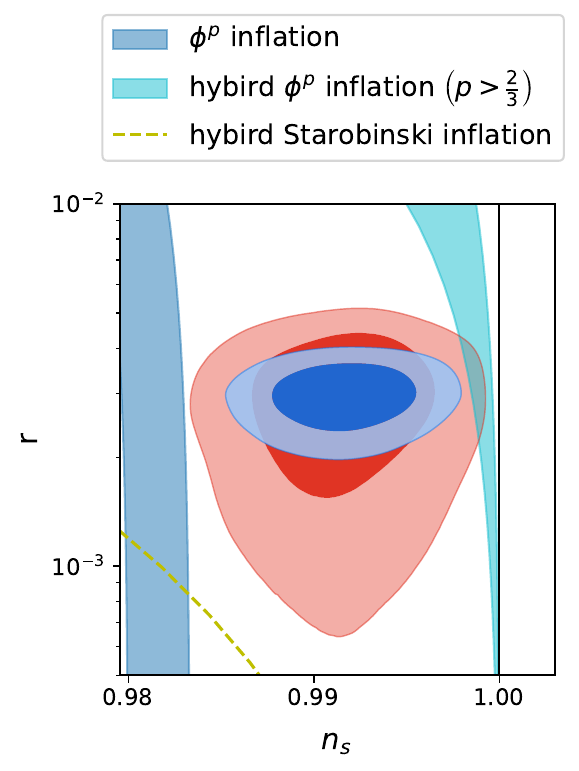}
    \end{subfigure}
\caption{The forecasted $r$-$n_s$ contours with CMB-S4 and CMB
S4+LiteBIRD in AdSEDE. Based on the bestfit results of AdSEDE for PR4+BK18,
we fix $r$ to 0 (left) and 0.003 (right). We also plot the
predictions of some inflation models on $n_s$ and $r$, the same as
\autoref{fig:ns_r}.}
    \label{fig:ns_r_future}
\end{figure}

\begin{table}[h]
    \centering
    \begin{tabular}{|c|c|} \hline \hline
        parameters & values \\ \hline
        $\log_{10}(z_c)            $ &  $3.512$\\
        $f_\text{EDE}              $ &  $0.112$\\
        $\log(10^{10} A_\mathrm{s})$ &  $3.076$\\
        $n_\mathrm{s}   $ &  $0.9916$\\
        $H_0            $ &  $71.92$\\
        $\Omega_\mathrm{b} h^2$ &  $0.02317$\\
        $\Omega_\mathrm{c} h^2$ &  $0.1362$\\
        $\tau_\mathrm{reio}$ & 0.0578 \\ \hline \hline
    \end{tabular}
    \caption{The fiducial values for the cosmological parameters used to create the forecast.}
    \label{tab:fiducial}
\end{table}

{
Here, we assume AdS-EDE as the fiducial cosmological model.
And set the cosmological parameters to the best-fit values
\footnote{When searching for the best-fit values, we fixed $r=0$ as we expect that $r \lesssim 0.003$ will not have a significant impact on the other parameters.}
based on the PR4+BK18 dataset.
The detailed values are shown in \autoref{tab:fiducial}.
We then considered different $r$ as the fiducial values.
The results are shown in \autoref{fig:ns_r_future},
where we set $r=0$ (left) and $r=0.003$ (right) respectively.
}
The $\phi^p$ inflation ($p>2/3$) and
the power-law inflation with a hybrid end will be ruled out at
$2\sigma$ level, if $r$ is still undetected, while the Starobinski
inflation with $N_*\thickapprox 300$ is consistent, since
\begin{equation} r={12\over N_*^2}\sim {\cal O}(1/10^4).
\end{equation} However, if $r=0.003$,
which would be detected by CMB-S4 and LiteBIRD, the $\phi^p$
inflation ($p>2/3$), power-law inflation and Starobinski inflation
all will be ruled out at $2\sigma$ level, only the $\phi^p$
inflation with $p<2/3$ with a hybrid end might survive.


\section{Conclusion}
\label{sec:conclusion}

It has been widely thought that the conflicts in cosmological
observations imply modifications beyond $\Lambda$CDM model.
However, such modifications, specially the injection of EDE before
the recombination, might be bringing a unforeseen impact on
searching for primordial GW and setting the value of $n_s$, so our
perspective on inflation.

Here, with the latest CMB datasets, we found for EDE that
$|n_s-1|\lesssim {\cal O}(0.01)$, while the upper limit of $r$ is
also slightly tighten, $r<0.028$ with Planck PR4+BK18 and $r<0.030$
with Planck PR4+ACT+SPT+BK18 dataset, which is consistent with the
results with Planck PR3+BK18 \cite{Ye:2022afu}. In light of our
constraint on $r-n_s$, the inflation models allowed by the results
in the $\Lambda$CDM model, such as Starobinski inflation, will be
excluded. However, in corresponding models satisfying (\ref{ns}),
if inflation ends by a waterfall instability when inflaton is
still at a deep slow-roll region, $n_s$ can be lifted close to
$n_s= 1$, so that the inflation models which have been ruled out,
such as the $\phi^p$ inflation, and also the Starobinski model
become possible again with their hybrid variants. It is also
interesting to explore other models with $|n_s-1|\lesssim {\cal
O}(0.01)$.

In upcoming decade, the combination of the
CMB-S4\cite{CMB-S4:2016ple} and the LiteBIRD
satellite\cite{LiteBIRD:2020khw} will be able to reach
$\sigma(n_s) \sim 0.005$ and $\sigma(r) < 10^{-3}$. Here, we also
show that the different inflation models allowed by the present
observations (if we happen to live with EDE) would be
distinguished by both experiments.

{ 
Here, we only force the Hubble tension in our analysis.
However, it is important to note that there are also many inconsistencies in other cosmological observations, such as $S_8$ tension.
Some of them call for the modification to the $\Lambda$CDM model (see e.g.\cite{Perivolaropoulos:2021jda,Abdalla:2022yfr} for reviews),
and may also affect the $r-n_s$ contour in \autoref{fig:ns_r}, which will impact our perspective on the inflation models.
Therefore, it is worth questioning the relevant issues beyond just the Hubble tension.
}

\acknowledgments

This work is supported by the NSFC No.12075246, the Fundamental
Research Funds for the Central Universities.

\appendix

\section{Priors for cosmological parameters} \label{sec:priors}

\begin{table}[H]
\centering
\begin{tabular}{|c|c|} \hline \hline
parameters          & priors         \\ \hline
$r$                 & $[0, 0.5]$     \\
$\ln(10^{10}A_s)$   & $[1.61, 3.91]$ \\
$n_s$               & $[0.8, 1.2]$   \\
$H_0$               & $[20, 100]$    \\
$\omega_b$          & $[0.005, 0.1]$ \\
$\omega_\text{cdm}$ & $[0.05, 0.99]$ \\
$\tau_\text{reio}$  & $[0.01, 0.8]$  \\ \hline
$\log_{10}(1+z_c)$  & $[2, 4.5]$     \\
$f_\text{EDE}$      & $[0, 0.3]$     \\
$\Theta_\mathrm{i}$ & $[0, 3.1]$     \\ \hline
$n_t$               & $[-10, 10]$    \\
$\alpha_s$          & $[-0.1, 0.1]$  \\ \hline \hline
\end{tabular}
\caption{Priors for cosmological parameters used in this work. All of them are flat priors for corresponding parameters. $\log_{10}(1+z_c)$ and $f_\text{EDE}$ are only used for EDE models. $\Theta_\mathrm{i}$ are only used for the axion-like EDE model. $n_s$ and $\alpha_s$ are only used in \autoref{sec:appendix}.}
\end{table}

\section{Results for $n_t$ and $\alpha_s$}
\label{sec:appendix}

\begin{figure}[h]
    \centering
    \includegraphics{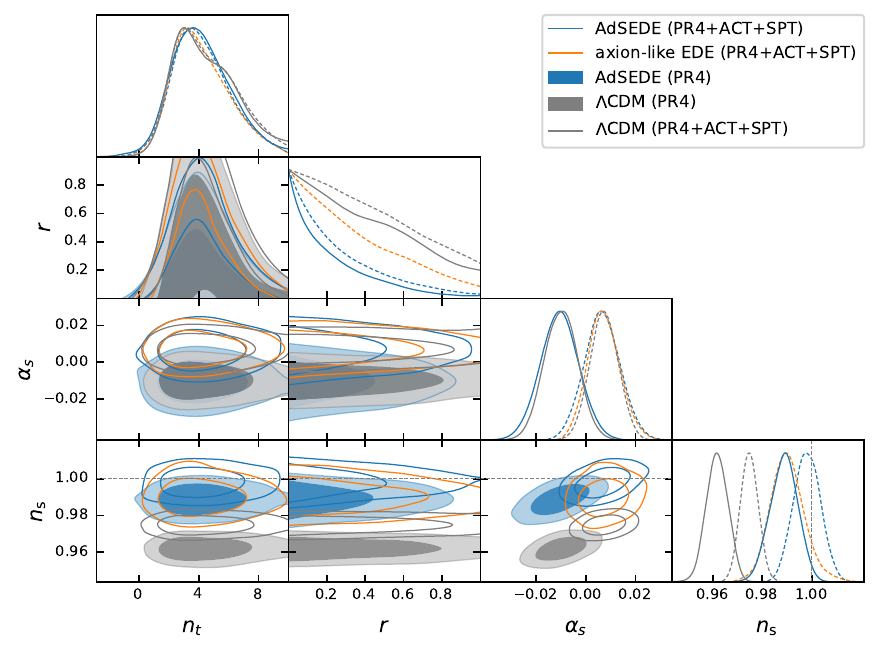}
\caption{Marginalized posterior distributions (68\% and 95\%
confidence intervals) for relevant parameters under different
models and datasets. BK18 is included in all the dataset.}
    \label{fig:nt_as}
\end{figure}

In \autoref{fig:nt_as}, we present the result for the spectral
tilt $n_t$ of primordial GW and the running of scalar spectral
index $\alpha_s = \mathrm{d}n_s/\mathrm{d}k$. We imposed the flat
priors on both $n_t$ and $\alpha_s$. We did not find the
significant effects of EDE on the observations of $n_t$ and
$\alpha_s$. The main difference on $\alpha_s$ is attributed to the
different combination of CMB datasets.

\begin{figure}[h]
    \centering
    \includegraphics{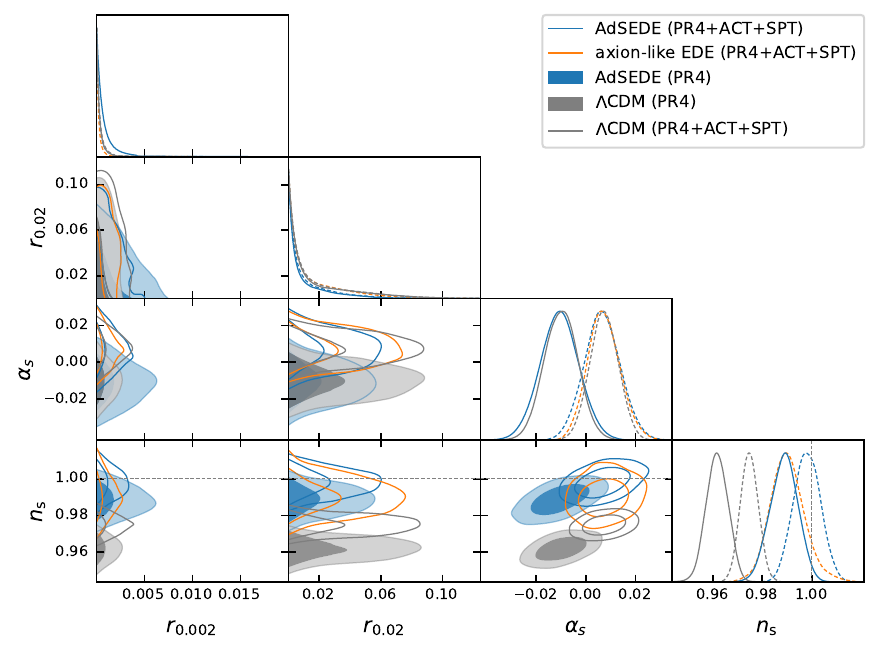}
\caption{Marginalized posterior distributions (68\% and 95\%
confidence intervals) for relevant parameters under different
models and datasets. BK18 is included from all the dataset.}
    \label{fig:r1_r2}
\end{figure}

However, when we switch from ($r_\text{pivot}, n_t$) to ($r_{k_1},
r_{k_2}$) through
\begin{equation}
    r_k = r_\text{pivot} \left( \frac{k}{k_\text{pivot}} \right)^{n_t-n_s+1}
\end{equation}
with $k_1=0.002$Mpc$^{-1}$ and $k_1=0.02$Mpc$^{-1}$, in light of
Planck18 \cite{Planck:2018jri}. We find their discrepancy on
$r_2$, as shown in \autoref{fig:r1_r2}, which is constrained to
$r_2 < 0.0457$ and $r_2 < 0.0566$ (95\% CL) for AdSEDE and
axion-like EDE, respectively, with the PR4+ACT+SPT CMB dataset
while $r_2<0.0424$ (95\% CL) for AdSEDE with the PR4 CMB dataset.
Their values are lower than $r_2 < 0.0653$ (95\% CL) for
$\Lambda$CDM with the PR4 CMB dataset
and $r_2 < 0.0656$ (95\% CL) with the PR4+ACT+SPT CMB dataset.
These discrepancy is more significant than that of
$r_\text{pivot}$. The main reason is that in EDE the lensing
spectrum at this scale ($\ell \sim 250$) is larger
\cite{Ye:2022afu}.

\section{On noise power spectrum and delensing}
\label{sec:appendixB}

In \autoref{sec:fur_constraint} we investigate the impact of EDE
on the constraining power of CMB-S4 and LiteBIRD. The noise power
spectrum $N_{\ell}$ for CMB-S4 is taken from the wiki of CMB-S4.
\footnote{\url{https://cmb-s4.uchicago.edu/wiki/index.php/Survey_Performance_Expectations}}
And for LiteBIRD, we consider a noise curve of
\begin{equation}
    N_{\ell}^{X Y}=s^2 \exp \left(\ell(\ell+1) \frac{\theta_{\text{FWHM}}^2}{8 \log
    2}\right),
\end{equation}
where the temperature noise is $s=2 \mu$K-arcmin,
$\theta_{\text{FWHM}} = 30$ arcmin for the full-width half-maximum
beam size \cite{LiteBIRD:2022cnt}, and the polarization noise has
additional factor $\sqrt{2}$.

Delensing on the CMB maps can help improve constraints on $r$
as well as reduce the effects of the cosmological models on the
lensing, specially the EDE model.
We simply model it as
\begin{equation}
    C_\ell = A_L C_\ell^\text{lensed} + (1-A_L) C_\ell^\text{unlensed}
\end{equation}
for the $C_\ell$ in \autoref{eq:Fisher}.
The delensing efficiency factor $A_L=0.27$ is
considered for CMB-S4
\footnote{\url{https://cmb-s4.uchicago.edu/wiki/index.php/Estimates_of_delensing_efficiency}}
and $A_L=0.57$ for LiteBIRD \cite{LiteBIRD:2022cnt}.  In addition, we model the
effects of thermal dust \cite{Planck:2014dmk} and synchrotron
\cite{Choi:2015xha} for those polarisation noise power spectrums which have not yet taken the foreground into account as
\begin{equation} A_{\text {dust }} \ell^{-2.42},\quad\quad
A_{\text {synch }} \ell^{-2.3},\end{equation} respectively, where
$A_{\text {dust }} $ and $A_{\text {synch }}$ will be regarded as
the nuisance parameters and be marginalised away.

\section{Marginalized posterior distributions for all cosmological parameters}
\label{sec:all}
\begin{figure}[H]
    \centering
    \includegraphics[width=\linewidth]{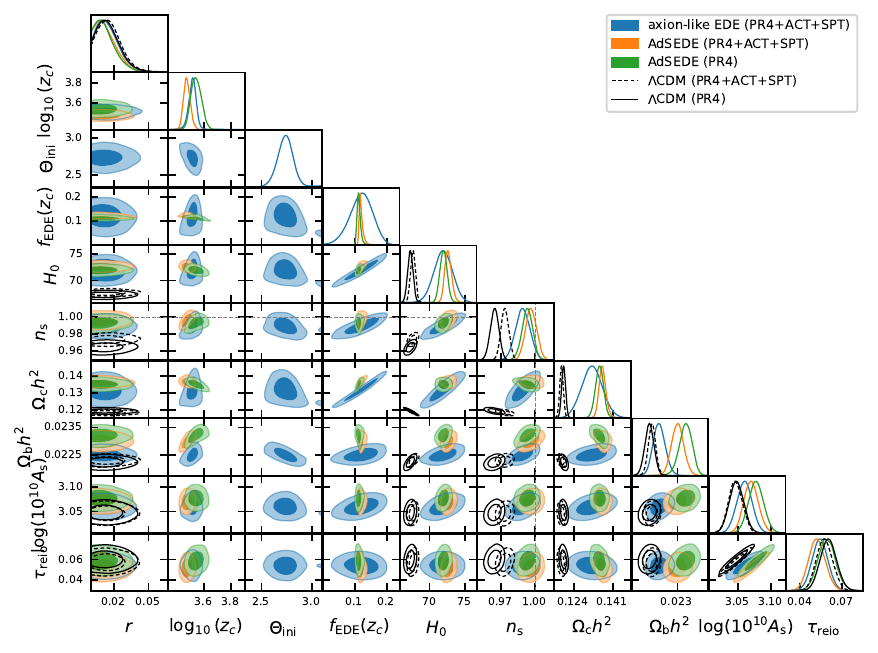}
\caption{Marginalized posterior distributions (68\% and 95\%
confidence intervals) for all cosmological parameters in different
models with different datasets. BK18 is included in all datasets.
Constraint of R21\cite{Riess:2021jrx} on $H_0$ is shown as a gray band.}
    \label{fig:EDE_constraints_all}
\end{figure}

\begin{table}[H]
    \centering
    \scriptsize
    \begin{tabular}{c|ccc} \hline \hline
parameters & $\Lambda$CDM (PR4+ACT+SPT) & $\Lambda$CDM (PR4) \\ \hline
$r              $ & $<0.0352$ & $r<0.0330$ \\
$\log(10^{10} A_\mathrm{s})$ &  $3.046(3.0376)\pm 0.011    $& $3.048(3.0424)\pm0.011$ \\
$n_\mathrm{s}   $ &  $0.9736(0.97236)\pm 0.0034 $& $0.9645(0.96627)\pm0.0038$ \\
$H_0            $ &  $67.68(67.546)\pm 0.34$& $67.31(67.395)\pm0.38$ \\
$\Omega_\mathrm{b} h^2$ &  $0.02231(0.022304)\pm 0.00010$& $0.02224(0.022257)\pm0.00012$ \\
$\Omega_\mathrm{c} h^2$ &  $0.11865(0.11887)\pm 0.00079$& $0.11956(0.11928)\pm 0.00084$ \\
$\tau_\mathrm{reio}$ &  $0.0575(0.0515)\pm 0.0057  $& $0.0603(0.0575)\pm 0.0059$ \\ \hline
$\Omega_\mathrm{m}         $ &  $0.3092(0.31084)\pm 0.0047 $& $0.3144(0.3130) \pm 0.0051$ \\
$S_8                       $ &  $0.8220(0.8209)\pm 0.0090    $& $0.8302(0.8256)\pm 0.0095$ \\
\hline \hline
    \end{tabular}
\caption{68\% confidence intervals for the parameters in $\Lambda$CDM model with different datasets, the best-fit values are shown
in parentheses. And for the one-tailed distribution we show 95\%
confidence intervals. BK18 is included in all datasets.}
    \label{tab:LCDM_constraints}
\end{table}


\bibliographystyle{JHEP}
\bibliography{./refs.bib}

\providecommand{\href}[2]{#2}\begingroup\raggedright\begin{thebibliography}{10}

\bibitem{Planck:2018vyg}
{\bf Planck} Collaboration, N.~Aghanim et~al., {\it {Planck 2018 results. VI.
  Cosmological parameters}},  {\em Astron. Astrophys.} {\bf 641} (2020) A6,
  [\href{http://arxiv.org/abs/1807.06209}{{\tt arXiv:1807.06209}}]. [Erratum:
  Astron.Astrophys. 652, C4 (2021)].

\bibitem{BICEP:2021xfz}
{\bf BICEP, Keck} Collaboration, P.~A.~R. Ade et~al., {\it {Improved
  Constraints on Primordial Gravitational Waves using Planck, WMAP, and
  BICEP/Keck Observations through the 2018 Observing Season}},  {\em Phys. Rev.
  Lett.} {\bf 127} (2021), no.~15 151301,
  [\href{http://arxiv.org/abs/2110.00483}{{\tt arXiv:2110.00483}}].

\bibitem{Perivolaropoulos:2021jda}
L.~Perivolaropoulos and F.~Skara, {\it {Challenges for \ensuremath{\Lambda}CDM:
  An update}},  {\em New Astron. Rev.} {\bf 95} (2022) 101659,
  [\href{http://arxiv.org/abs/2105.05208}{{\tt arXiv:2105.05208}}].

\bibitem{Abdalla:2022yfr}
E.~Abdalla et~al., {\it {Cosmology intertwined: A review of the particle
  physics, astrophysics, and cosmology associated with the cosmological
  tensions and anomalies}},  {\em JHEAp} {\bf 34} (2022) 49--211,
  [\href{http://arxiv.org/abs/2203.06142}{{\tt arXiv:2203.06142}}].

\bibitem{DiValentino:2019qzk}
E.~Di~Valentino, A.~Melchiorri, and J.~Silk, {\it {Planck evidence for a closed
  Universe and a possible crisis for cosmology}},  {\em Nature Astron.} {\bf 4}
  (2019), no.~2 196--203, [\href{http://arxiv.org/abs/1911.02087}{{\tt
  arXiv:1911.02087}}].

\bibitem{Handley:2019tkm}
W.~Handley, {\it {Curvature tension: evidence for a closed universe}},  {\em
  Phys. Rev. D} {\bf 103} (2021), no.~4 L041301,
  [\href{http://arxiv.org/abs/1908.09139}{{\tt arXiv:1908.09139}}].

\bibitem{DiValentino:2016hlg}
E.~Di~Valentino, A.~Melchiorri, and J.~Silk, {\it {Reconciling Planck with the
  local value of $H_0$ in extended parameter space}},  {\em Phys. Lett. B} {\bf
  761} (2016) 242--246, [\href{http://arxiv.org/abs/1606.00634}{{\tt
  arXiv:1606.00634}}].

\bibitem{Mortsell:2018mfj}
E.~M\"ortsell and S.~Dhawan, {\it {Does the Hubble constant tension call for
  new physics?}},  {\em JCAP} {\bf 09} (2018) 025,
  [\href{http://arxiv.org/abs/1801.07260}{{\tt arXiv:1801.07260}}].

\bibitem{Vagnozzi:2019ezj}
S.~Vagnozzi, {\it {New physics in light of the $H_0$ tension: An alternative
  view}},  {\em Phys. Rev. D} {\bf 102} (2020), no.~2 023518,
  [\href{http://arxiv.org/abs/1907.07569}{{\tt arXiv:1907.07569}}].

\bibitem{Knox:2019rjx}
L.~Knox and M.~Millea, {\it {Hubble constant hunter\textquoteright{}s guide}},
  {\em Phys. Rev. D} {\bf 101} (2020), no.~4 043533,
  [\href{http://arxiv.org/abs/1908.03663}{{\tt arXiv:1908.03663}}].

\bibitem{DiValentino:2019jae}
E.~Di~Valentino, A.~Melchiorri, O.~Mena, and S.~Vagnozzi, {\it {Nonminimal dark
  sector physics and cosmological tensions}},  {\em Phys. Rev. D} {\bf 101}
  (2020), no.~6 063502, [\href{http://arxiv.org/abs/1910.09853}{{\tt
  arXiv:1910.09853}}].

\bibitem{Schoneberg:2021qvd}
N.~Sch\"oneberg, G.~Franco~Abell\'an, A.~P\'erez~S\'anchez, S.~J. Witte,
  V.~Poulin, and J.~Lesgourgues, {\it {The H0 Olympics: A fair ranking of
  proposed models}},  {\em Phys. Rept.} {\bf 984} (2022) 1--55,
  [\href{http://arxiv.org/abs/2107.10291}{{\tt arXiv:2107.10291}}].

\bibitem{Karwal:2016vyq}
T.~Karwal and M.~Kamionkowski, {\it {Dark energy at early times, the Hubble
  parameter, and the string axiverse}},  {\em Phys. Rev. D} {\bf 94} (2016),
  no.~10 103523, [\href{http://arxiv.org/abs/1608.01309}{{\tt
  arXiv:1608.01309}}].

\bibitem{Poulin:2018cxd}
V.~Poulin, T.~L. Smith, T.~Karwal, and M.~Kamionkowski, {\it {Early Dark Energy
  Can Resolve The Hubble Tension}},  {\em Phys. Rev. Lett.} {\bf 122} (2019),
  no.~22 221301, [\href{http://arxiv.org/abs/1811.04083}{{\tt
  arXiv:1811.04083}}].

\bibitem{Ye:2021nej}
G.~Ye, B.~Hu, and Y.-S. Piao, {\it {Implication of the Hubble tension for the
  primordial Universe in light of recent cosmological data}},  {\em Phys. Rev.
  D} {\bf 104} (2021), no.~6 063510,
  [\href{http://arxiv.org/abs/2103.09729}{{\tt arXiv:2103.09729}}].

\bibitem{Jiang:2022uyg}
J.-Q. Jiang and Y.-S. Piao, {\it {Toward early dark energy and ns=1 with
  Planck, ACT, and SPT observations}},  {\em Phys. Rev. D} {\bf 105} (2022),
  no.~10 103514, [\href{http://arxiv.org/abs/2202.13379}{{\tt
  arXiv:2202.13379}}].

\bibitem{Smith:2022hwi}
T.~L. Smith, M.~Lucca, V.~Poulin, G.~F. Abellan, L.~Balkenhol, K.~Benabed,
  S.~Galli, and R.~Murgia, {\it {Hints of early dark energy in Planck, SPT, and
  ACT data: New physics or systematics?}},  {\em Phys. Rev. D} {\bf 106}
  (2022), no.~4 043526, [\href{http://arxiv.org/abs/2202.09379}{{\tt
  arXiv:2202.09379}}].

\bibitem{Jiang:2022qlj}
J.-Q. Jiang, G.~Ye, and Y.-S. Piao, {\it {Return of Harrison-Zeldovich spectrum
  in light of recent cosmological tensions}},
  \href{http://arxiv.org/abs/2210.06125}{{\tt arXiv:2210.06125}}.

\bibitem{Cruz:2022oqk}
J.~S. Cruz, F.~Niedermann, and M.~S. Sloth, {\it {A grounded perspective on new
  early dark energy using ACT, SPT, and BICEP/Keck}},  {\em JCAP} {\bf 02}
  (2023) 041, [\href{http://arxiv.org/abs/2209.02708}{{\tt arXiv:2209.02708}}].

\bibitem{Ye:2020btb}
G.~Ye and Y.-S. Piao, {\it {Is the Hubble tension a hint of AdS phase around
  recombination?}},  {\em Phys. Rev. D} {\bf 101} (2020), no.~8 083507,
  [\href{http://arxiv.org/abs/2001.02451}{{\tt arXiv:2001.02451}}].

\bibitem{Ye:2022afu}
G.~Ye and Y.-S. Piao, {\it {Improved constraints on primordial gravitational
  waves in light of the H0 tension and BICEP/Keck data}},  {\em Phys. Rev. D}
  {\bf 106} (2022), no.~4 043536, [\href{http://arxiv.org/abs/2202.10055}{{\tt
  arXiv:2202.10055}}].

\bibitem{DiValentino:2018zjj}
E.~Di~Valentino, A.~Melchiorri, Y.~Fantaye, and A.~Heavens, {\it {Bayesian
  evidence against the Harrison-Zel\textquoteright{}dovich spectrum in tensions
  with cosmological data sets}},  {\em Phys. Rev. D} {\bf 98} (2018), no.~6
  063508, [\href{http://arxiv.org/abs/1808.09201}{{\tt arXiv:1808.09201}}].

\bibitem{Giare:2022rvg}
W.~Giar\`e, F.~Renzi, O.~Mena, E.~Di~Valentino, and A.~Melchiorri, {\it {Is the
  Harrison-Zel\textquoteright{}dovich spectrum coming back? ACT preference for
  ns \ensuremath{\sim} 1 and its discordance with Planck}},  {\em Mon. Not.
  Roy. Astron. Soc.} {\bf 521} (2023), no.~2 2911--2918,
  [\href{http://arxiv.org/abs/2210.09018}{{\tt arXiv:2210.09018}}].

\bibitem{Calderon:2023obf}
R.~Calder\'on, A.~Shafieloo, D.~K. Hazra, and W.~Sohn, {\it {On the consistency
  of \ensuremath{\Lambda}CDM with CMB measurements in light of the latest
  Planck, ACT and SPT data}},  {\em JCAP} {\bf 08} (2023) 059,
  [\href{http://arxiv.org/abs/2302.14300}{{\tt arXiv:2302.14300}}].

\bibitem{Planck:2020olo}
{\bf Planck} Collaboration, Y.~Akrami et~al., {\it {$Planck$ intermediate
  results. LVII. Joint Planck LFI and HFI data processing}},  {\em Astron.
  Astrophys.} {\bf 643} (2020) A42,
  [\href{http://arxiv.org/abs/2007.04997}{{\tt arXiv:2007.04997}}].

\bibitem{ACT:2020frw}
{\bf ACT} Collaboration, S.~K. Choi et~al., {\it {The Atacama Cosmology
  Telescope: a measurement of the Cosmic Microwave Background power spectra at
  98 and 150 GHz}},  {\em JCAP} {\bf 12} (2020) 045,
  [\href{http://arxiv.org/abs/2007.07289}{{\tt arXiv:2007.07289}}].

\bibitem{SPT-3G:2021eoc}
{\bf SPT-3G} Collaboration, D.~Dutcher et~al., {\it {Measurements of the E-mode
  polarization and temperature-E-mode correlation of the CMB from SPT-3G 2018
  data}},  {\em Phys. Rev. D} {\bf 104} (2021), no.~2 022003,
  [\href{http://arxiv.org/abs/2101.01684}{{\tt arXiv:2101.01684}}].

\bibitem{SPT-3G:2022hvq}
{\bf SPT-3G} Collaboration, L.~Balkenhol et~al., {\it {Measurement of the CMB
  temperature power spectrum and constraints on cosmology from the SPT-3G 2018
  TT, TE, and EE dataset}},  {\em Phys. Rev. D} {\bf 108} (2023), no.~2 023510,
  [\href{http://arxiv.org/abs/2212.05642}{{\tt arXiv:2212.05642}}].

\bibitem{SimonsObservatory:2018koc}
{\bf Simons Observatory} Collaboration, P.~Ade et~al., {\it {The Simons
  Observatory: Science goals and forecasts}},  {\em JCAP} {\bf 02} (2019) 056,
  [\href{http://arxiv.org/abs/1808.07445}{{\tt arXiv:1808.07445}}].

\bibitem{CMB-S4:2016ple}
{\bf CMB-S4} Collaboration, K.~N. Abazajian et~al., {\it {CMB-S4 Science Book,
  First Edition}},  \href{http://arxiv.org/abs/1610.02743}{{\tt
  arXiv:1610.02743}}.

\bibitem{LiteBIRD:2020khw}
{\bf LiteBIRD} Collaboration, M.~Hazumi et~al., {\it {LiteBIRD: JAXA's new
  strategic L-class mission for all-sky surveys of cosmic microwave background
  polarization}},  {\em Proc. SPIE Int. Soc. Opt. Eng.} {\bf 11443} (2020)
  114432F, [\href{http://arxiv.org/abs/2101.12449}{{\tt arXiv:2101.12449}}].

\bibitem{Couchot:2016vaq}
F.~Couchot, S.~Henrot-Versill\'e, O.~Perdereau, S.~Plaszczynski,
  B.~Rouill\'e~d'Orfeuil, M.~Spinelli, and M.~Tristram, {\it {Cosmology with
  the cosmic microwave background temperature-polarization correlation}},  {\em
  Astron. Astrophys.} {\bf 602} (2017) A41,
  [\href{http://arxiv.org/abs/1609.09730}{{\tt arXiv:1609.09730}}].

\bibitem{Tristram:2020wbi}
M.~Tristram et~al., {\it {Planck constraints on the tensor-to-scalar ratio}},
  {\em Astron. Astrophys.} {\bf 647} (2021) A128,
  [\href{http://arxiv.org/abs/2010.01139}{{\tt arXiv:2010.01139}}].

\bibitem{Planck:2019nip}
{\bf Planck} Collaboration, N.~Aghanim et~al., {\it {Planck 2018 results. V.
  CMB power spectra and likelihoods}},  {\em Astron. Astrophys.} {\bf 641}
  (2020) A5, [\href{http://arxiv.org/abs/1907.12875}{{\tt arXiv:1907.12875}}].

\bibitem{Carron:2022eyg}
J.~Carron, M.~Mirmelstein, and A.~Lewis, {\it {CMB lensing from Planck
  PR4~maps}},  {\em JCAP} {\bf 09} (2022) 039,
  [\href{http://arxiv.org/abs/2206.07773}{{\tt arXiv:2206.07773}}].

\bibitem{Beutler:2011hx}
F.~Beutler, C.~Blake, M.~Colless, D.~H. Jones, L.~Staveley-Smith, L.~Campbell,
  Q.~Parker, W.~Saunders, and F.~Watson, {\it {The 6dF Galaxy Survey: Baryon
  Acoustic Oscillations and the Local Hubble Constant}},  {\em Mon. Not. Roy.
  Astron. Soc.} {\bf 416} (2011) 3017--3032,
  [\href{http://arxiv.org/abs/1106.3366}{{\tt arXiv:1106.3366}}].

\bibitem{Ross:2014qpa}
A.~J. Ross, L.~Samushia, C.~Howlett, W.~J. Percival, A.~Burden, and M.~Manera,
  {\it {The clustering of the SDSS DR7 main Galaxy sample \textendash{} I. A 4
  per cent distance measure at $z = 0.15$}},  {\em Mon. Not. Roy. Astron. Soc.}
  {\bf 449} (2015), no.~1 835--847, [\href{http://arxiv.org/abs/1409.3242}{{\tt
  arXiv:1409.3242}}].

\bibitem{eBOSS:2020yzd}
{\bf eBOSS} Collaboration, S.~Alam et~al., {\it {Completed SDSS-IV extended
  Baryon Oscillation Spectroscopic Survey: Cosmological implications from two
  decades of spectroscopic surveys at the Apache Point Observatory}},  {\em
  Phys. Rev. D} {\bf 103} (2021), no.~8 083533,
  [\href{http://arxiv.org/abs/2007.08991}{{\tt arXiv:2007.08991}}].

\bibitem{Cuceu:2019for}
A.~Cuceu, J.~Farr, P.~Lemos, and A.~Font-Ribera, {\it {Baryon Acoustic
  Oscillations and the Hubble Constant: Past, Present and Future}},  {\em JCAP}
  {\bf 10} (2019) 044, [\href{http://arxiv.org/abs/1906.11628}{{\tt
  arXiv:1906.11628}}].

\bibitem{Schoneberg:2022ggi}
N.~Sch\"oneberg, L.~Verde, H.~Gil-Mar\'\i{}n, and S.~Brieden, {\it {BAO+BBN
  revisited \textemdash{} growing the Hubble tension with a 0.7 km/s/Mpc
  constraint}},  {\em JCAP} {\bf 11} (2022) 039,
  [\href{http://arxiv.org/abs/2209.14330}{{\tt arXiv:2209.14330}}].

\bibitem{BOSS:2016wmc}
{\bf BOSS} Collaboration, S.~Alam et~al., {\it {The clustering of galaxies in
  the completed SDSS-III Baryon Oscillation Spectroscopic Survey: cosmological
  analysis of the DR12 galaxy sample}},  {\em Mon. Not. Roy. Astron. Soc.} {\bf
  470} (2017), no.~3 2617--2652, [\href{http://arxiv.org/abs/1607.03155}{{\tt
  arXiv:1607.03155}}].

\bibitem{Brout:2022vxf}
D.~Brout et~al., {\it {The Pantheon+ Analysis: Cosmological Constraints}},
  {\em Astrophys. J.} {\bf 938} (2022), no.~2 110,
  [\href{http://arxiv.org/abs/2202.04077}{{\tt arXiv:2202.04077}}].

\bibitem{Addison:2015wyg}
G.~E. Addison, Y.~Huang, D.~J. Watts, C.~L. Bennett, M.~Halpern, G.~Hinshaw,
  and J.~L. Weiland, {\it {Quantifying discordance in the 2015 Planck CMB
  spectrum}},  {\em Astrophys. J.} {\bf 818} (2016), no.~2 132,
  [\href{http://arxiv.org/abs/1511.00055}{{\tt arXiv:1511.00055}}].

\bibitem{Planck:2016tof}
{\bf Planck} Collaboration, N.~Aghanim et~al., {\it {Planck intermediate
  results. LI. Features in the cosmic microwave background temperature power
  spectrum and shifts in cosmological parameters}},  {\em Astron. Astrophys.}
  {\bf 607} (2017) A95, [\href{http://arxiv.org/abs/1608.02487}{{\tt
  arXiv:1608.02487}}].

\bibitem{Motloch:2019gux}
P.~Motloch and W.~Hu, {\it {Lensinglike tensions in the $Planck$ legacy
  release}},  {\em Phys. Rev. D} {\bf 101} (2020), no.~8 083515,
  [\href{http://arxiv.org/abs/1912.06601}{{\tt arXiv:1912.06601}}].

\bibitem{McDonough:2022pku}
E.~McDonough and M.~Scalisi, {\it {Towards Early Dark Energy in string
  theory}},  {\em JHEP} {\bf 10} (2023) 118,
  [\href{http://arxiv.org/abs/2209.00011}{{\tt arXiv:2209.00011}}].

\bibitem{Cicoli:2023qri}
M.~Cicoli, M.~Licheri, R.~Mahanta, E.~McDonough, F.~G. Pedro, and M.~Scalisi,
  {\it {Early Dark Energy in Type IIB String Theory}},  {\em JHEP} {\bf 06}
  (2023) 052, [\href{http://arxiv.org/abs/2303.03414}{{\tt arXiv:2303.03414}}].

\bibitem{Ye:2020oix}
G.~Ye and Y.-S. Piao, {\it {$T_0$ censorship of early dark energy and AdS
  vacua}},  {\em Phys. Rev. D} {\bf 102} (2020), no.~8 083523,
  [\href{http://arxiv.org/abs/2008.10832}{{\tt arXiv:2008.10832}}].

\bibitem{Jiang:2021bab}
J.-Q. Jiang and Y.-S. Piao, {\it {Testing AdS early dark energy with Planck,
  SPTpol, and LSS data}},  {\em Phys. Rev. D} {\bf 104} (2021), no.~10 103524,
  [\href{http://arxiv.org/abs/2107.07128}{{\tt arXiv:2107.07128}}].

\bibitem{Torrado:2020dgo}
J.~Torrado and A.~Lewis, {\it {Cobaya: Code for Bayesian Analysis of
  hierarchical physical models}},  {\em JCAP} {\bf 05} (2021) 057,
  [\href{http://arxiv.org/abs/2005.05290}{{\tt arXiv:2005.05290}}].

\bibitem{Blas:2011rf}
D.~Blas, J.~Lesgourgues, and T.~Tram, {\it {The Cosmic Linear Anisotropy
  Solving System (CLASS) II: Approximation schemes}},  {\em JCAP} {\bf 07}
  (2011) 034, [\href{http://arxiv.org/abs/1104.2933}{{\tt arXiv:1104.2933}}].

\bibitem{Riess:2021jrx}
A.~G. Riess et~al., {\it {A Comprehensive Measurement of the Local Value of the
  Hubble Constant with 1 km s$^{-1}$ Mpc$^{-1}$ Uncertainty from the Hubble
  Space Telescope and the SH0ES Team}},  {\em Astrophys. J. Lett.} {\bf 934}
  (2022), no.~1 L7, [\href{http://arxiv.org/abs/2112.04510}{{\tt
  arXiv:2112.04510}}].

\bibitem{LaPosta:2021pgm}
A.~La~Posta, T.~Louis, X.~Garrido, and J.~C. Hill, {\it {Constraints on
  prerecombination early dark energy from SPT-3G public data}},  {\em Phys.
  Rev. D} {\bf 105} (2022), no.~8 083519,
  [\href{http://arxiv.org/abs/2112.10754}{{\tt arXiv:2112.10754}}].

\bibitem{Tristram:2021tvh}
M.~Tristram et~al., {\it {Improved limits on the tensor-to-scalar ratio using
  BICEP and Planck data}},  {\em Phys. Rev. D} {\bf 105} (2022), no.~8 083524,
  [\href{http://arxiv.org/abs/2112.07961}{{\tt arXiv:2112.07961}}].

\bibitem{Starobinsky:1980te}
A.~A. Starobinsky, {\it {A New Type of Isotropic Cosmological Models Without
  Singularity}},  {\em Phys. Lett. B} {\bf 91} (1980) 99--102.

\bibitem{Silverstein:2008sg}
E.~Silverstein and A.~Westphal, {\it {Monodromy in the CMB: Gravity Waves and
  String Inflation}},  {\em Phys. Rev. D} {\bf 78} (2008) 106003,
  [\href{http://arxiv.org/abs/0803.3085}{{\tt arXiv:0803.3085}}].

\bibitem{McAllister:2008hb}
L.~McAllister, E.~Silverstein, and A.~Westphal, {\it {Gravity Waves and Linear
  Inflation from Axion Monodromy}},  {\em Phys. Rev. D} {\bf 82} (2010) 046003,
  [\href{http://arxiv.org/abs/0808.0706}{{\tt arXiv:0808.0706}}].

\bibitem{Abbott:1984fp}
L.~F. Abbott and M.~B. Wise, {\it {Constraints on Generalized Inflationary
  Cosmologies}},  {\em Nucl. Phys. B} {\bf 244} (1984) 541--548.

\bibitem{Mukhanov:2013tua}
V.~Mukhanov, {\it {Quantum Cosmological Perturbations: Predictions and
  Observations}},  {\em Eur. Phys. J. C} {\bf 73} (2013) 2486,
  [\href{http://arxiv.org/abs/1303.3925}{{\tt arXiv:1303.3925}}].

\bibitem{Kallosh:2013hoa}
R.~Kallosh and A.~Linde, {\it {Universality Class in Conformal Inflation}},
  {\em JCAP} {\bf 07} (2013) 002, [\href{http://arxiv.org/abs/1306.5220}{{\tt
  arXiv:1306.5220}}].

\bibitem{Roest:2013fha}
D.~Roest, {\it {Universality classes of inflation}},  {\em JCAP} {\bf 01}
  (2014) 007, [\href{http://arxiv.org/abs/1309.1285}{{\tt arXiv:1309.1285}}].

\bibitem{Martin:2013tda}
J.~Martin, C.~Ringeval, and V.~Vennin, {\it {Encyclop\ae{}dia Inflationaris}},
  {\em Phys. Dark Univ.} {\bf 5-6} (2014) 75--235,
  [\href{http://arxiv.org/abs/1303.3787}{{\tt arXiv:1303.3787}}].

\bibitem{Kallosh:2022ggf}
R.~Kallosh and A.~Linde, {\it {Hybrid cosmological attractors}},  {\em Phys.
  Rev. D} {\bf 106} (2022), no.~2 023522,
  [\href{http://arxiv.org/abs/2204.02425}{{\tt arXiv:2204.02425}}].

\bibitem{Ye:2022efx}
G.~Ye, J.-Q. Jiang, and Y.-S. Piao, {\it {Toward inflation with ns=1 in light
  of the Hubble tension and implications for primordial gravitational waves}},
  {\em Phys. Rev. D} {\bf 106} (2022), no.~10 103528,
  [\href{http://arxiv.org/abs/2205.02478}{{\tt arXiv:2205.02478}}].

\bibitem{Linde:1983gd}
A.~D. Linde, {\it {Chaotic Inflation}},  {\em Phys. Lett. B} {\bf 129} (1983)
  177--181.

\bibitem{Kaloper:2008fb}
N.~Kaloper and L.~Sorbo, {\it {A Natural Framework for Chaotic Inflation}},
  {\em Phys. Rev. Lett.} {\bf 102} (2009) 121301,
  [\href{http://arxiv.org/abs/0811.1989}{{\tt arXiv:0811.1989}}].

\bibitem{Linde:1991km}
A.~D. Linde, {\it {Axions in inflationary cosmology}},  {\em Phys. Lett. B}
  {\bf 259} (1991) 38--47.

\bibitem{Linde:1993cn}
A.~D. Linde, {\it {Hybrid inflation}},  {\em Phys. Rev. D} {\bf 49} (1994)
  748--754, [\href{http://arxiv.org/abs/astro-ph/9307002}{{\tt
  astro-ph/9307002}}].

\bibitem{DAmico:2020euu}
G.~D'Amico and N.~Kaloper, {\it {Rollercoaster cosmology}},  {\em JCAP} {\bf
  08} (2021) 058, [\href{http://arxiv.org/abs/2011.09489}{{\tt
  arXiv:2011.09489}}].

\bibitem{DAmico:2021vka}
G.~D'Amico, N.~Kaloper, and A.~Westphal, {\it {Double Monodromy Inflation: A
  Gravity Waves Factory for CMB-S4, LiteBIRD and LISA}},  {\em Phys. Rev. D}
  {\bf 104} (2021), no.~8 L081302, [\href{http://arxiv.org/abs/2101.05861}{{\tt
  arXiv:2101.05861}}].

\bibitem{Takahashi:2021bti}
F.~Takahashi and W.~Yin, {\it {Cosmological implications of ns
  \ensuremath{\approx} 1 in light of the Hubble tension}},  {\em Phys. Lett. B}
  {\bf 830} (2022) 137143, [\href{http://arxiv.org/abs/2112.06710}{{\tt
  arXiv:2112.06710}}].

\bibitem{DAmico:2021fhz}
G.~D'Amico, N.~Kaloper, and A.~Westphal, {\it {General double monodromy
  inflation}},  {\em Phys. Rev. D} {\bf 105} (2022), no.~10 103527,
  [\href{http://arxiv.org/abs/2112.13861}{{\tt arXiv:2112.13861}}].

\bibitem{Abazajian:2019eic}
K.~Abazajian et~al., {\it {CMB-S4 Science Case, Reference Design, and Project
  Plan}},  \href{http://arxiv.org/abs/1907.04473}{{\tt arXiv:1907.04473}}.

\bibitem{Jungman:1995av}
G.~Jungman, M.~Kamionkowski, A.~Kosowsky, and D.~N. Spergel, {\it {Weighing the
  universe with the cosmic microwave background}},  {\em Phys. Rev. Lett.} {\bf
  76} (1996) 1007--1010, [\href{http://arxiv.org/abs/astro-ph/9507080}{{\tt
  astro-ph/9507080}}].

\bibitem{Planck:2018jri}
{\bf Planck} Collaboration, Y.~Akrami et~al., {\it {Planck 2018 results. X.
  Constraints on inflation}},  {\em Astron. Astrophys.} {\bf 641} (2020) A10,
  [\href{http://arxiv.org/abs/1807.06211}{{\tt arXiv:1807.06211}}].

\bibitem{LiteBIRD:2022cnt}
{\bf LiteBIRD} Collaboration, E.~Allys et~al., {\it {Probing Cosmic Inflation
  with the LiteBIRD Cosmic Microwave Background Polarization Survey}},  {\em
  PTEP} {\bf 2023} (2023), no.~4 042F01,
  [\href{http://arxiv.org/abs/2202.02773}{{\tt arXiv:2202.02773}}].

\bibitem{Planck:2014dmk}
{\bf Planck} Collaboration, R.~Adam et~al., {\it {Planck intermediate results.
  XXX. The angular power spectrum of polarized dust emission at intermediate
  and high Galactic latitudes}},  {\em Astron. Astrophys.} {\bf 586} (2016)
  A133, [\href{http://arxiv.org/abs/1409.5738}{{\tt arXiv:1409.5738}}].

\bibitem{Choi:2015xha}
S.~K. Choi and L.~A. Page, {\it {Polarized galactic synchrotron and dust
  emission and their correlation}},  {\em JCAP} {\bf 12} (2015) 020,
  [\href{http://arxiv.org/abs/1509.05934}{{\tt arXiv:1509.05934}}].

\end{thebibliography}\endgroup
\end{document}